\documentclass[letterpaper, 10 pt, conference]{ieeeconf}  

\IEEEoverridecommandlockouts                              

\usepackage[
placement=top,
angle=0,
color=red,
scale=1.1,
hshift=0,
vshift=-15
]{background}
\backgroundsetup{contents=\bf{----------------------------------------------------------------- PREPRINT -----------------------------------------------------------------}}

\title{\LARGE \bf
Bearing-based Target Localisation in Search and Rescue Scenarios
}

\author{Giulia Michieletto$^{1,3}$, Nicola Mimmo$^2$, Roberto Naldi$^2$, and Angelo Cenedese$^1$ 
\thanks{*This work has been supported by the Grant “Design Of Cooperative Energy-aware Aerial plaTforms for remote and contact-aware operations” (DOCEAT) funded by the Italian MUR, n. 2020RTWES4, CUP n. E63C22000410001.}
\thanks{$^1$ G. Michieletto is with the Dep. of Management and Engineering, University of Padova, Italy.}
\thanks{$^2$ N. Mimmo and R. Naldi are with the
	Det. of Electrical, Electronic and Information Engineering "G. Marconi", University of Bologna, Italy.}
        \thanks{$^3$ A. Cenedese and G. Michieletto are with the Dep. of Information Engineering, University of Padova, Italy.} 
\thanks{Corresponding contact: {\tt\small giulia.michieletto@unipd.it}}
}

\usepackage{amsfonts,amsmath}
\usepackage{tikz}
\usepackage{xcolor,graphicx}
\usepackage{caption,subcaption}

\usepackage{stfloats}

\newcommand{\R}{\mathbb{R}}
\newcommand{\N}{\mathbb{N}}

\newcommand{\F}{\mathcal{F}}

\DeclareMathOperator{\col}{col}

\newtheorem{hyp}{Assumption}
\newtheorem{prob}{Problem}
\newtheorem{rem}{Remark}

\begin{document}

\maketitle

\begin{abstract}
This paper deals with the target localisation problem in search and rescue scenarios in which the technology is based on electromagnetic transceivers. The noise floor and the shape of the electromagnetic radiation pattern make this problem challenging. 
Indeed, on the one hand, the signal-to-noise ratio reduces with the inverse of the distance from the electromagnetic source thus impacting estimation-based techniques applicability. On the other hand, non-isotropic radiation patterns lessen the efficacy of gradient-based policies. 
In this work, we manage a fleet of autonomous agents, equipped with electromagnetic sensors, by combining gradient-based and estimation-based techniques to speed up the transmitter localisation. Simulations specialized in the ARTVA technology used in search and rescue in avalanche scenarios confirm that our scheme outperforms current solutions.
\end{abstract}

\section{Introduction}

{Target localisation is subject to time constraints, especially in Search and Rescue (SAR) operations in critical environments such as avalanches~\cite{falk1994avalanche}}. In these contexts,  deploying a fleet of autonomous agents with strategic sensing and actuation capabilities can streamline the localisation effort. 
Hence, coordinating multi-agent systems to collaboratively localise a target, while simultaneously addressing estimation and control problems, remains an active research area~\cite{Javaid2023Communication,Mu2023UAV}.
Particularly challenging are the SAR scenarios in which the technology is based on electromagnetic transceivers due to the noise floor and the shape of the radiation pattern.

In general, most of the existing solutions for target localisation through multi-agent systems adopt a double-layer architecture, comprising a target position estimator and an agents' trajectory planner, commonly leveraging the concept of \textit{active sensing}. 
In more detail, as for the target position estimate, the state-of-art literature can be classified into two main categories based on the nature of inferred information: direction-related and distance-related~\cite{Alhafnawi2023Survey}. Distance-related schemes access direct or indirect distance measurements. In contrast, algorithms based on direction-related data rely on bearing information, which consists of a unitary-length vector pointing to the target. 
When accounting for \textit{bearing-only} data, the estimation task is commonly addressed through two primary approaches. The first relies on traditional and evolutionary Kalman filtering techniques, and the second utilizes batch-form methods based on ordinary Least Squares (LS)~\cite{zou2022target}. 
The LS-based approaches are generally more effective in handling the non-ideal nature of the measurements compared to Kalman-based methods, which often struggle with instability induced by data noise~\cite{aidala1979kalman}. 
 In the case of electromagnetic transceivers, the bearing uncertainty due to a small Signal-to-Noise Ratio (SNR) impacts also the LS-based estimation algorithms' efficacy especially when the sensing agents are far from the target representing the electromagnetic source. Since the SNR decreases with the distance, a suitable active sensing policy should steer the multi-agent system toward the estimated target. 

If the target represents a signal source, the Extremum Seeking (ES) represents a valid alternative to the traditional double-layer architecture. Indeed, ES adjusts the agents' position to maximize the received signal power through a gradient-based policy \cite{scheinker2024100}. In practice, the agents' movement induced by ES is twofold. On the one hand, the agents explore their neighbourhood to estimate the local gradient and, on the other hand, the agents exploit this estimation to move toward the source. A possible drawback of this technique is a slow convergence speed due to the inherent time scale separation between the exploration and exploitation phases. Moreover, since the target localisation is achieved by moving the agents to the source, non-isotropic radiation patterns make the gradient-based search pattern longer than necessary, further slowing down the localisation procedure.

In this work, we tackle the localisation of a stationary target consisting in an electromagnetic source by using a fleet of autonomous vehicles equipped with the so-called ARTVA technology~\cite{Silvagni2017Multipurpose}, usually adopted in avalanche scenarios. In particular, we combine an ES scheme with a centralized bearing-only LS-based estimation technique to speed up the localisation procedure. In detail, each agent implements an ES scheme to estimate the local gradient, whose normalization is interpreted as a bearing. Then, the localisation technique, inspired by~\cite{pozzan2023active}, exploits the estimated bearings through a recursive LS paradigm. Finally, the multi-agent system is regulated by combining ES-based search movements and formation-to-target estimate distance minimization. 
The integration of these two approaches is dynamically regulated depending on the quality of the gathered bearings.

Through numerical simulations, we show the advantages of steering the centroid of a multi-agent system through a combination of ES and LS target position estimates. In particular, we show that the localisation time of our novel approach is shorter than those of individual ES and LS target localisation algorithms.

\paragraph*{Notation}
We use $\R$ and $\N$ to describe the set of real numbers and the set of natural numbers greater than zero, respectively. 
The set of rotation matrices is represented by $\text{SO}(3)$, and if $R \in \text{SO}(3)$, then $R R^\top = I$ and $\det(R)=1$. 
The symbol $\langle \cdot, \cdot \rangle$ denotes the scalar product. Let $X_i \in \R^{n_i \times n}$ for $i \in \{ 1 \dots m\}$, with $n_i , n, m \in \mathbb{N}$, be matrices. Then, we define $\col(X_1,\ldots,X_m)$ as the column stack of $X_1$, $\dots$, $X_m$.

\begin{figure*}[t!]
    \centering
        \begin{minipage}[b]{0.56\textwidth}
        \centering
\includegraphics[width=0.98\columnwidth]{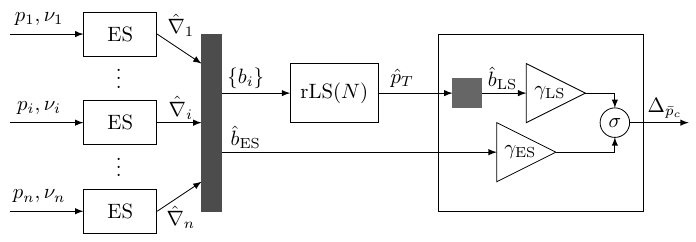}
    \caption{Scheme of the proposed solution involving $n$ ARTVA's gradient estimators based on ES paradigm, a recursive LS centralized TX position estimator, and a combinatorial formation controller.}
    \label{fig:scheme}
    \end{minipage}
    \hfill
    \begin{minipage}[b]{0.4\textwidth}
    \centering
\begin{tikzpicture}
\begin{scope}[scale=0.8]
\def\sFx{0.90};
\def\sFy{0.5};
    \draw[->] ({0.5*\sFx},0) -- ({8.5*\sFx},0) node[right]{$k$};
    \draw[->] ({0.5*\sFx},0) -- ({0.5*\sFx},{5*\sFy}) node[right]{$t,\tau$};
    \draw[dotted, line width=0.075mm] ({8.5*\sFx},{1*\sFy}) -- ({0.5*\sFx},{1*\sFy}) node[left]{$1$};
    \draw[dotted, line width=0.075mm] ({8.5*\sFx},{2*\sFy}) -- ({0.5*\sFx},{2*\sFy}) node[left]{$2$};
    \draw[dotted, line width=0.075mm] ({8.5*\sFx},{3*\sFy}) -- ({0.5*\sFx},{3*\sFy}) node[left, yshift = 3]{$\vdots$};
    \draw[dotted, line width=0.075mm] ({8.5*\sFx},{4*\sFy}) -- ({0.5*\sFx},{4*\sFy}) node[left]{$N$};
    
    \draw[dotted, line width=0.075mm] ({1*\sFx},{5*\sFy}) -- ({1*\sFx},{0*\sFy}) node[below]{$1$};
    \draw[dotted, line width=0.075mm] ({2*\sFx},{5*\sFy}) -- ({2*\sFx},{0*\sFy}) node[below]{$2$};
    \draw[dotted, line width=0.075mm] ({3*\sFx},{5*\sFy}) -- ({3*\sFx},{0*\sFy}) node[below, yshift = -2]{$\cdots$};
    \draw[dotted, line width=0.075mm] ({4*\sFx},{5*\sFy}) -- ({4*\sFx},{0*\sFy}) node[below]{$N$};
    \draw[dotted, line width=0.075mm] ({5*\sFx},{5*\sFy}) -- ({5*\sFx},{0*\sFy}) node[below]{$N\!\!+\!\!1$};
    \draw[dotted, line width=0.075mm] ({6*\sFx},{5*\sFy}) -- ({6*\sFx},{0*\sFy}) node[below]{$N\!\!+\!\!2$};
    \draw[dotted, line width=0.075mm] ({7*\sFx},{5*\sFy}) -- ({7*\sFx},{0*\sFy}) node[below, yshift = -2]{$\cdots$};
    \draw[dotted, line width=0.075mm] ({8*\sFx},{5*\sFy}) -- ({8*\sFx},{0*\sFy}) node[below]{$2N$};
    
    \draw ({1*\sFx},{1*\sFy}) circle[radius=3pt];
    \draw ({2*\sFx},{1*\sFy}) circle[radius=3pt];
    \node[] at ({3*\sFx},{1*\sFy}) {$\cdots$};
    \draw ({4*\sFx},{1*\sFy}) circle[radius=3pt];
    
    \fill ({1*\sFx},{1*\sFy}) circle[radius=2pt];
    \fill ({2*\sFx},{2*\sFy}) circle[radius=2pt];
    \node[rotate = 70] at ({3*\sFx-0.1},{3*\sFy}) {$\ddots$};
    \fill ({4*\sFx},{4*\sFy}) circle[radius=2pt];

    \draw ({5*\sFx},{2*\sFy}) circle[radius=3pt];
    \draw ({6*\sFx},{2*\sFy}) circle[radius=3pt];
    \node[] at ({7*\sFx},{2*\sFy}) {$\cdots$};
    \draw ({8*\sFx},{2*\sFy}) circle[radius=3pt];

    \fill ({5*\sFx},{1*\sFy}) circle[radius=2pt];
    \fill ({6*\sFx},{2*\sFy}) circle[radius=2pt];
    \node[rotate = 70] at ({7*\sFx-0.1},{3*\sFy}) {$\ddots$};    
    \fill ({8*\sFx},{4*\sFy}) circle[radius=2pt];


    \end{scope}
\end{tikzpicture}
    \caption{Behavior of the counters $t$ (white circles) and $\tau$ (black dots) which, every $N$ steps, are incremented and reset, respectively.}
    \label{fig:time}
    \end{minipage}
\end{figure*}
\section{Problem Statement and Proposed Solution}

\subsection{Sensing Technology}
\label{sec:SensTech}
The developed automatic search technique relies on the ARTVA system, composed of devices operating in two modalities, \textit{i.e.}, transmitter (TX), and receiver (RX). Roughly, the ARTVA technology relies on electromagnetic fields emitted by coils of wires wrapped around ferromagnetic cores. 

We introduce the reference frames $\F_I$, $\F_T$, and $\F_{i}$, with $i \in \{1 \ldots n \}$ and $n \in \mathbb{N}$, to model the system. Frame $\F_I$ represents an inertial reference, $\F_T$ is rigidly attached to the TX device (assumed unique) and $\F_{i}$ are rigidly attached to the $i$th RX device. Let the inertial position of the TX device be $p_T  \in \R^3$, and the inertial position of the $i$th RX device be $p_i  \in \R^3$. Moreover, let $R_T  \in \text{SO}(3)$ be a rotation matrix describing the attitude of $\F_T$ to $\F_I$. Then, as reported in \cite{Pinies2006Fast}, the magnetic field generated by TX evaluated at position $p_i $, and expressed in $\F_i$, is 
\begin{equation}
\nonumber
    \label{eq:H}
    h_i(p_i ) = \dfrac{\Pi}{4\pi}  R_i R_T^\top  \dfrac{m(R_T(p_i -p_T ))}{\|p_i -p_T \|^5}
\end{equation}
where $\Pi >0$ denotes the nominal power associated with TX and $m(p) := \col(2 x^2-y^2-z^2,\, 3 x y,\,3 x z)$, with $p := \col(x,y,z)$ and $x,y,x \in \R$. In this paper, we assume the RXs are equipped with isometric ARTVA antennae \cite{Lavorgna2007Multi}, whereas the TX's signal is anisotropic.
Therefore, assuming the antennae are subject to a uniform noise, we model the output of the $i$th RX as
\begin{equation}
\nonumber
\label{eq:v}
    v(p_i ,\nu_i) = h_i(p_i )+\nu_i
\end{equation}
with $\nu_i \in \R^3$. 
From this, we compute the signal intensity received by the $i$th RX device as
\begin{equation}
    \label{eq:y}
   y(p_i ,\nu_i)  = \| v(p_i ,\nu_i) \|{^{-2/3}}
\end{equation}
{where the power $-2/3$ has been introduced to ensure $y(p_i,0) \propto \|p_i-p_T\|^2$, $y(\cdot,0) \in \mathcal{C}^2$, and $y(p_T ,\cdot) = 0$. }

\subsection{Problem Statement}

With the quantities described in {Section} \ref{sec:SensTech} at hand, we now introduce the standing assumption framing the problem.

\begin{hyp}[Noise Ergodicity \& Boundedness]
\label{hyp:noise} The noise $\nu_i(t)$ is a position-independent zero-mean ergodic stochastic process and $\exists\,  \overline{\nu}\,:\, \|\nu_i(t)\|_\infty \le \overline{\nu}, \forall i = 1,\ldots, n$.	
\end{hyp}

We formally state the following problem.

\begin{prob}
\label{prob}
    Design a centralised algorithm with inputs $p_{i}$, and $y(p_i,\nu_i)$, with {$i = 1, \dots, n$}, such that:
    \begin{enumerate}
        \item there exist two non-empty sets $\mathcal{P}_0 \subseteq \R^3$ and $\hat{\mathcal{P}}_0 \subseteq \R^3$ such that the trajectories of the RXs and the TX position estimator, $p_i(t)$ and  $\hat{p}_T(t)$, remain bounded for any $t > 0$, for any initial condition $(p_{i0},\hat{p}_{T0}) \in \mathcal{P}_0\times \hat{\mathcal{P}}_0$, with $i = 1, \dots, n$;
        \item for every $\epsilon \ge 0$  there exists $t^\star \ge 0$ such that $\limsup_{t \to \infty} \|\hat{p}_T(t)-p_T\| < \epsilon$, 
for $t \ge t^\star$.  
    \end{enumerate}
\end{prob}

\begin{rem}
    Solving this problem represents a hard task. Indeed,  the inverse of the distance law embedded into~\eqref{eq:y} makes the SNR go to zero for $\|p_i-p_T\|\to \infty$ and vice-versa, to infinite when $\|p_i-p_T\|\to 0$. 
\end{rem}

\subsection{Proposed Solution}
\label{sec:ProposedSolution}

The proposed solution to {Problem}~\ref{prob} consists of three components: an ARTVA's gradient estimator, a target's position estimator, and a formation controller ({Figure}~\ref{fig:scheme}). In more detail, each RX runs a private ARTVA's gradient estimator, which elaborates $y(p_i,\nu_i)$ and $p_i$ and provides $\hat{\nabla}_i \in \mathbb{R}^3$ representing an estimation of $\nabla_{p_i} y(p_i,\nu_i)$. 
All $\hat{\nabla}_i$ feed a centralised algorithm in charge of estimating the TX position, namely $\hat{p}_T$. Finally, the formation control consists of a centralised algorithm that moves the formation at any research step through a convex union of the estimated directions $\hat{\nabla}_i$ and $\hat{p}_T-p_i$, for $i = 1,\ldots, n$. 

The solution is described as a discrete-time system in which we use two counters, namely $t$ and $\tau$, evolving as periodic functions of the step-index $k \in\N$. Let $N \in \N$ be the desired period. Then, we introduce
\begin{equation}
    \label{eq:time}
    \begin{aligned}
    \tau(k+1) =& \left\{\!\!\!\begin{array}{cc}
    1 & k /N= \lfloor k/N \rfloor\\
    \tau(k)+1 &  \text{otherwise}
    \end{array}\right. \\
    t(k) =& \,\lceil k/N \rceil \\
    \end{aligned}
\end{equation}
with $\tau(1) = 1$, and where $\lfloor \cdot \rfloor$ and $\lceil \cdot \rceil$ are the floor and the ceiling operators. 
In essence, we reset $\tau$ and increment $t$ every $N$ steps ({Figure}~\ref{fig:time}). The scheme allows the TX position estimator to run at every step, resetting it every $N$ steps. Conversely, {the gradient estimator} and formation controller {run once every $N$ steps}. Hence, the gradient estimator and formation controller evolve with $t$, while the TX position estimator operates on $\tau$. 
Intuitively, we update the formation only at the end of each optimisation step.

\subsubsection{Gradient Estimator}

For estimating $\nabla_{p_i} y(p_i,\nu_i)$ we exploit an ES approach. In detail, the ES algorithm perturbs the RX's position by forcing $p_i$ to follow a reference trajectory. Then, we correlate the measurements $y(p_i,\nu_i)$ with the reference trajectory to provide, at the average, $\hat{\nabla}_i$. 
Technically, we define the controlled position $p_i$ of the $i$th RX as the sum of two components, \textit{i.e.}, a base position $\bar{p}_i$ which will be updated by the formation controller, and a dither trajectory $\delta \Gamma(\omega t)$ with amplitude $\delta >0$, and pulsation $\omega>0$ and where $\Gamma(\omega t):=\col(\sin(\omega t),\cos(\omega t), \sin({\kappa} \omega t))${, with $\kappa \ge 0$}. The position is then defined as
\begin{subequations}
\label{eq:ES}
\begin{equation}
    \label{eq:pi}
    p_i(t) = \bar{p}_i(t) + \delta \Gamma(\omega t). 
\end{equation}
As for the gradient estimator, we propose 
    \begin{align}
    \label{eq:hat_nabla}
        \hat{\nabla}_i(t+1) =&\, (1-\alpha)\hat{\nabla}_i(t)+\alpha \varepsilon_i(t) \Gamma(\omega t)\\
        \label{eq:z}
        z_i(t+1)=&\,  z_i(t)+\alpha \varepsilon_i(t)\\
        \label{eq:varepsilon}
         \varepsilon_i(t) = &\, y(p_i(t) ,\nu_i(t))-z_i(t)
           \end{align}
\end{subequations}
with {initial conditions} $\hat{\nabla}_i(1) = \hat{\nabla}_{i1} \in \R^3$ {and} $ z_i(1) = z_1 \in \R$, and {where} $\alpha >0$ is a tunable parameter that will be designed later. Basically, \eqref{eq:hat_nabla} represents a low-pass filter of {$\varepsilon_i(t) \Gamma(\omega t)$} which, inspired by an ES approach, represents an instantaneous gradient estimation. It is worth noting that $\|\hat{\nabla}_i-\nabla_{p_i} y(p_i,\nu_i)\|$ is tunable through the design of $\delta$, $\omega$, and $\alpha$. 
The dynamics {\eqref{eq:z}-\eqref{eq:varepsilon}, \textit{i.e.},} $z_i$ with output $\varepsilon_i$, represent a high-pass filter whose goal is to get rid of the continuous component of $y(p_i(t) ,\nu_i(t))$. 

\subsubsection{TX position estimation}

The centralised TX position estimator collects $p_i{(t)}$ and $\hat{\nabla}_i{(t)}$, with $i = 1,\ldots, n$, and provides $\hat{p}_T{(t)}$. 
In more detail, we define the $i$th estimated bearing as the normalisation of $\hat{\nabla}_i$, namely
\begin{subequations}
    \label{eq:TXPosEstim}
\begin{equation}
    \label{eq:Bearing}
    b_i(t) := \hat{\nabla}_i(t)\big/\sqrt{\epsilon +\|\hat{\nabla}_i(t)\|^2},
\end{equation}
where $\epsilon > 0$ makes $b_i$ well-posed for any $\|\hat{\nabla}_i\| \ge 0$, and we collect all the $b_i$, with $i = 1,\ldots, n$, in the vector $b(t) := \col(b_1(t),\ldots,b_n(t))$.  Roughly, the idea is finding $\hat{p}_T$ that minimises $ \sum_i \| b_i-(\hat{p}_T-\bar{p}_i)/\|\hat{p}_T-\bar{p}_i\| \|^2$. {Intuitively,} in the lucky case that all $b_i$ are exactly oriented toward the TX, $\hat{p}_T$ could be computed as the point at the intersections of all $b_i$. At the opposite, when vectors $b_i$ are not oriented toward a common position, 
the minimisation could have no bounded solutions.  
We let 
\begin{equation}
\label{eq:barpc}
\bar{p}_c{(t)} :=  \dfrac{1}{n} \sum_{i = 1}^n \bar{p}_i{(t)}
\end{equation}
be the formation's centroid. Then, inspired by these arguments, we introduce $\hat{\rho}$ as a maximum norm for $\hat{p}_T-\bar{p}_c$ and $\mathcal{B}_{\hat{\rho}} \subset \R^3$ as a closed ball of radius $\hat{\rho}$.  Then, we denote with $\hat{p}_T^\star$ the solution to the following constrained minimisation
\begin{equation}
    \label{eq:hat_pt_star}
    \begin{aligned}
    \hat{p}_T^\star(t) :=&\, 
    {\arg\min}_{p} 
    \left\|b(t)-f(p,\bar{p}(t))\right\|^2
    \\
   \text{subject to}  &\, {{p}} \in \bar{p}_c(t)+\mathcal{B}_{\hat{\rho}}
    \end{aligned}
\end{equation}
in which $\bar{p}(t) = \col(\bar{p}_1(t), \,\ldots, \,\bar{p}_n(t))$, and 
\begin{equation}
\nonumber
    f(p,\bar{p}) = \col\left(\dfrac{p-\bar{p}_1}{\sqrt{\epsilon+\|p-\bar{p}_1\|^2}},\,\ldots,\,\dfrac{p-\bar{p}_n}{\sqrt{\epsilon+\|p-\bar{p}_n\|^2}}\right).
\end{equation}
\begin{figure*}[t]
     \centering
     \normalsize
\begin{equation}
\label{eq:hat_pt_plus}
\begin{aligned}
    \hat{p}_T({t,}\tau\!+\!1) =& \begin{cases}
        \bar{p}_c(t)+\dfrac{\hat{p}_T({t,}\tau)+ \Delta\hat{p}_T({t,}\tau)-\bar{p}_c(t)}{\|\hat{p}_T({t,}\tau)+ \Delta\hat{p}_T({t,}\tau)-\bar{p}_c(t)\|}\hat{\rho} & \|\hat{p}_T({t,}\tau)+ \Delta\hat{p}_T({t},\tau)-\bar{p}_c(t)\| > \hat{\rho} \\[0.3cm]
       \hat{p}_T({t,}\tau)+ \Delta\hat{p}_T({t,}\tau)  & \text{otherwise}
    \end{cases} 
    \\    \Delta\hat{p}_T({t,}\tau):=&\,\beta \left(F(\hat{p}_T({t,}\tau),\bar{p}(t))\right)^{\dag} \!\! (b(t)\!-\!f(\hat{p}_T({t,}\tau),\bar{p}(t))).
\end{aligned} 
\end{equation}
\hrule
 \end{figure*}

While the well-posedness of the problem \eqref{eq:hat_pt_star} is guaranteed by the constraints, its convexity depends on 
$\bar{p}_{c}$, and on $\hat{\nabla}_i$ via $b_i$. {Intuitively,} problem \eqref{eq:hat_pt_star} is {locally} convex when all the $\hat{\nabla}_i$ points inward the convex hull containing the formation.
Solutions to the problem \eqref{eq:hat_pt_star}, namely $\hat{p}_T$, are computed with a degree of approximation through recursive algorithms.  In particular, we propose to estimate the TX position through the recursive LS algorithm reported in~\eqref{eq:hat_pt_plus} at the {top} of the {next} page,  with initial conditions $\hat{p}_T({t},1)= \hat{p}_{T0}(t)$ and step size $\beta > 0$, where the superscript $\dag$ denotes the Moore-Penrose left pseudo-inverse,  and in which
%
    \begin{align}
        F(\hat{p}_T,\bar{p}) \!:=&\!\! \left[\!\!\!\begin{array}{c}
(\epsilon+\|\hat{p}_T-\bar{p}_1\|^2)^{-1/2}\Pr(\hat{p}_T-\bar{p}_1)\\
\vdots\\ 
(\epsilon+\|\hat{p}_T-\bar{p}_n\|^2)^{-1/2}\Pr(\hat{p}_T-\bar{p}_n)
        \end{array}\!\!\! \right]\\
        \nonumber
        \Pr(\hat{p}_T-\bar{p}_i) \! := &  I - (\epsilon+\|\hat{p}_T-\bar{p}_i\|^2)^{-1}
        (\hat{p}_T-\bar{p}_i)(\hat{p}_T-\bar{p}_i)^\top 
    \end{align}
%
with $i = 1,\ldots, n$.
 The term $F(\hat{p}_T,\bar{p}) = \left.\partial f(p,\bar{p})/\partial p\right|_{p = \hat{p}_T}$ represents the variation of $f$, evaluated at the current $\hat{p}_T$, induced by a variation of $\hat{p}_T$. Since $f$ is composed of normalised vectors, its partial derivative introduces the operator $(\epsilon+\|\hat{p}_T-\bar{p}_i\|^2)^{-1/2}\Pr(\hat{p}_T-\bar{p}_i)$ representing the projection of the variation of $\hat{p}_T-\bar{p}_i$ into the unitary sphere. 
 In practice, \eqref{eq:hat_pt_plus} looks for the $\hat{p}_T$, constrained into $\bar{p}_c+\mathcal{B}_{\hat{\rho}}$, making $\hat{p}_T-\bar{p}_i$ collinear to and oriented as $b_i$ for all $i = 1,\ldots, n$. 
  
 Finally, the initial condition $\hat{p}_{T0}(t)$ {represents an estimation reset} defined as
\begin{equation}
\nonumber
    \hat{p}_{T0}(t) := \arg\min_{p \in \R^3} \dfrac{1}{2} \sum_{i=1}^n (\bar{p}{_i}(t)-p)^\top \Pr(b{_i}(t))(\bar{p}{_i}(t)-p).
\end{equation}
Roughly, $\hat{p}_{T0}(t)$ is the solution to a point-to-line {distance minimisation} problem, whose solution is found in closed form by imposing zero gradient. It holds that
\begin{equation}
    \label{eq:hat_pt0}
 \!\hat{p}_{T0}(t) \!:= \! \!\left(\sum_{i=1}^n \Pr(b{_i}(t)) \right)^{-1} \!\!\sum_{i=1}^n \Pr(b{_i}(t))\bar{p}{_i}(t). 
\end{equation}
\end{subequations}
Note that {$\hat{p}_{T0}(t)$} is the point in $\mathbb{R}^3$ that minimizes the sum of distances to the parameterized lines $\ell_i(v):=\bar{p}{_i}(t)+b{_i}(t) v$, where $v \in \mathbb{R}$. In practice, {$\hat{p}_{T0}(t)$} typically lies near the intersection of all $\ell_i(v)$. On the one hand, {$\hat{p}_{T0}(t)$} is finite if the matrix $\left[b_1(t)\, \cdots\, b_n(t)\right]$ is full-rank but, on the other hand, this does not imply that $\langle {\hat{p}_{T0}(t)}-b_i(t), \,b_i(t) \rangle > 0$ for all $i = 1,\ldots, n$, \textit{i.e.}, the directions of ${\hat{p}_{T0}(t)}-b_i(t)$ and $b_i$ may disagree for some $i \in \{1, \dots, n\}$. This makes \eqref{eq:hat_pt0} a good reset for \eqref{eq:hat_pt_plus} but it cannot be used in place of \eqref{eq:hat_pt_plus}.

\subsubsection{Formation Controller}


{We outline a controller preserving the formation shape}. 
More in detail,  we define $p_{ib} \in \R^3$ as the $i$th agent's position relative to $\bar{p}_c$ such that
\begin{equation}
\nonumber
\bar{p}_i(t) := p_{ib} + \bar{p}_c(t).
\end{equation}
Then, we {impose} that $p_{ib}$ is a constant, for all {$i = 1, \dots, n$}.  As a consequence, we designed the formation controller for updating $\bar{p}_c$ to improve the accuracy of $\hat{p}_T$. In particular, we mixed, in a convex combination, the information coming from the ES, $b$, and that generated by the TX position estimator, $\hat{p}_T$. We combined these two via  {$\sigma \,:\,\mathbb{N} \times \mathbb{N} \to [0,\,1]$} such that $\sigma = 0$ means $\hat{p}_T$ is  error-free and $\sigma = 1$ means $\hat{p}_T$ is totally unreliable. Then, we defined 
\begin{subequations}
\label{eq:formation_controller}
    \begin{equation}
\label{eq:sigma}
    {\sigma({t,\tau})} := {\Vert (b(t)\!-\!f(\hat{p}_T({t, \tau}),\bar{p}(t))) \Vert}\big/({2 \sqrt{n}}) .
\end{equation}

The ES-based and LS-based directions toward which we move the formation are respectively defined as 
\begin{equation}
    \label{eq:mean_agents}
    \begin{aligned}
    \hat{b}_{\text{ES}}(t) :=&\,  \textstyle\sum_{i=1}^{n}  \hat{\nabla}_i(t)\bigg/ \sqrt{\epsilon+\left\|\textstyle\sum_{i=1}^{n}  \hat{\nabla}_i(t)\right\|^2} \\
        \hat{b}_{\text{LS}}({t, \tau}) := & \dfrac{(\hat{p}_T({t, \tau})-\bar{p}_c(t))}{\sqrt{\epsilon+\|\hat{p}_T({t, \tau})-\bar{p}_c(t)\|^2}}.
\end{aligned}
\end{equation}
Then, we propose the following policy to manage the fleet
\begin{equation}
\label{eq:bar_pc_plus}
\begin{aligned}
     \bar{p}_c(t+1) =&\, \bar{p}_c(t) + {\gamma\Big(}
     \sigma({t,N})  \hat{b}_{\text{ES}}(t)\\
    \quad & \!+\! 
    (1\!-\!\sigma({t,N}))\hat{b}_{\text{LS}}(t,N){ \Big)},
\end{aligned}
\end{equation}
\end{subequations}
where $\gamma>0$ is introduced to enforce time-scale separation between the fleet control and the target position estimation.

\begin{figure*}[t!]
    \centering
    \begin{subfigure}[b]{0.32\textwidth}
        \centering
        \includegraphics[width=\textwidth]{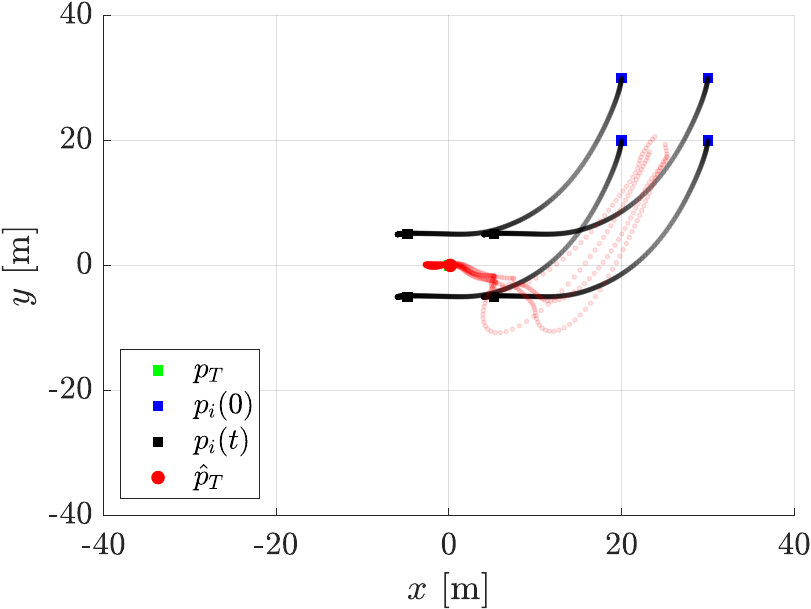}
        \caption*{}
    \end{subfigure}
    \hfill
    \begin{subfigure}[b]{0.32\textwidth}
        \centering
        \includegraphics[width=0.975\textwidth]{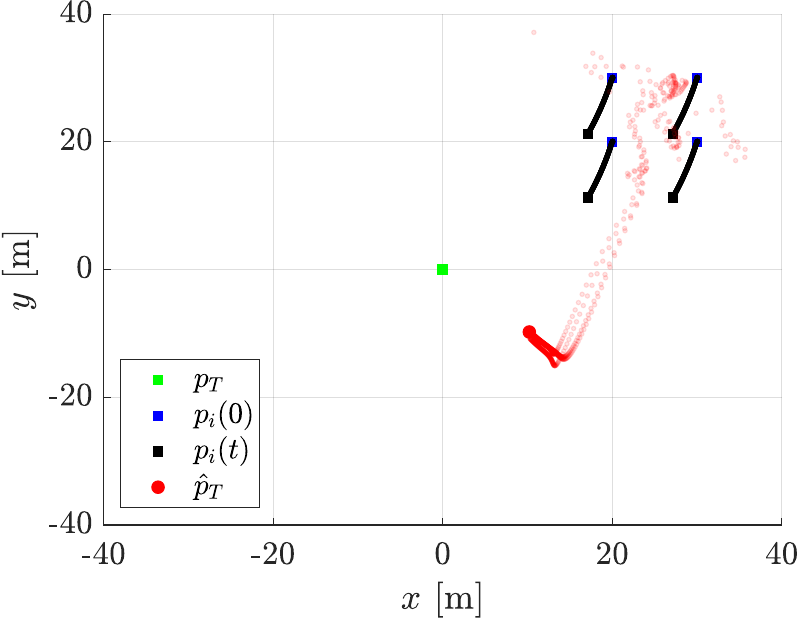}
        \caption*{}
    \end{subfigure}
    \hfill
    \begin{subfigure}[b]{0.32\textwidth}
        \centering
        \includegraphics[width=0.975\textwidth]{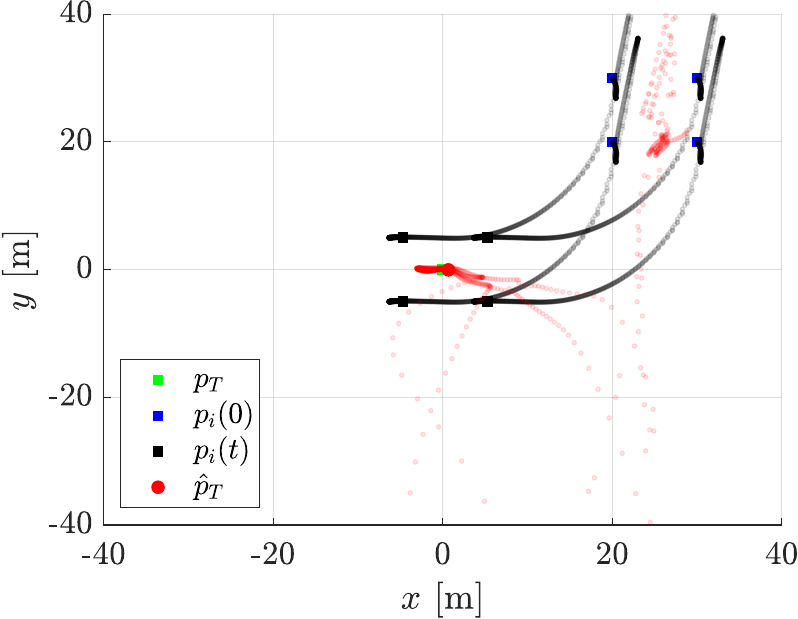}
        \caption*{}
    \end{subfigure}

   \vspace{-0.2cm} 
    
    \begin{subfigure}[b]{0.32\textwidth}
        \centering
        \includegraphics[width=\textwidth]{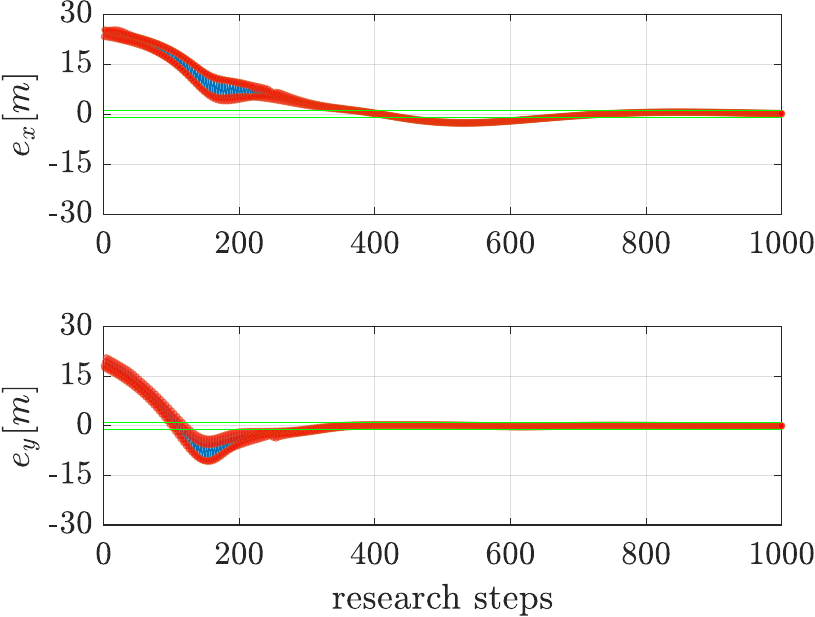}
        \caption{ES+LS}
        \label{fig:ES+LS}
    \end{subfigure}
    \hfill
    \begin{subfigure}[b]{0.32\textwidth}
        \centering
        \includegraphics[width=\textwidth]{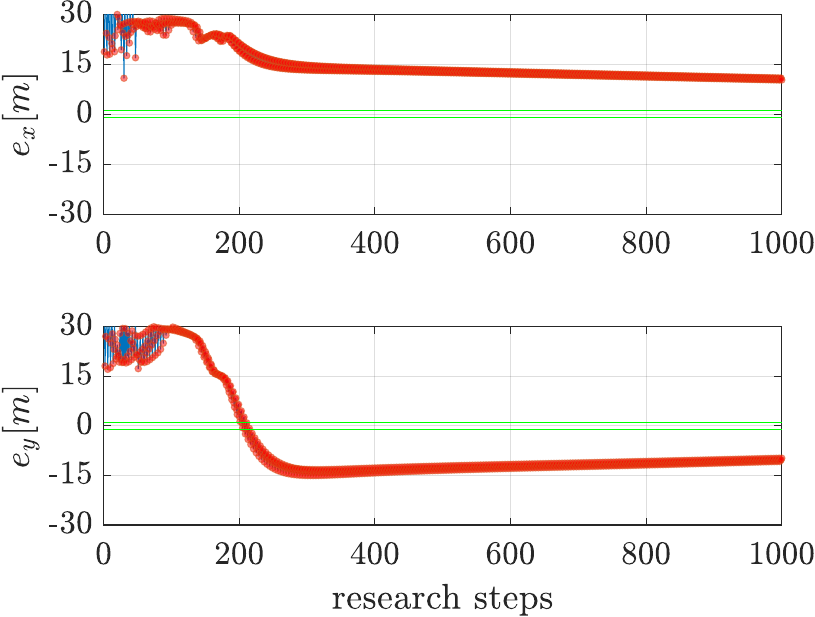}
        \caption{ES}
        \label{fig:ES}
    \end{subfigure}
    \hfill
    \begin{subfigure}[b]{0.32\textwidth}
        \centering
        \includegraphics[width=\textwidth]{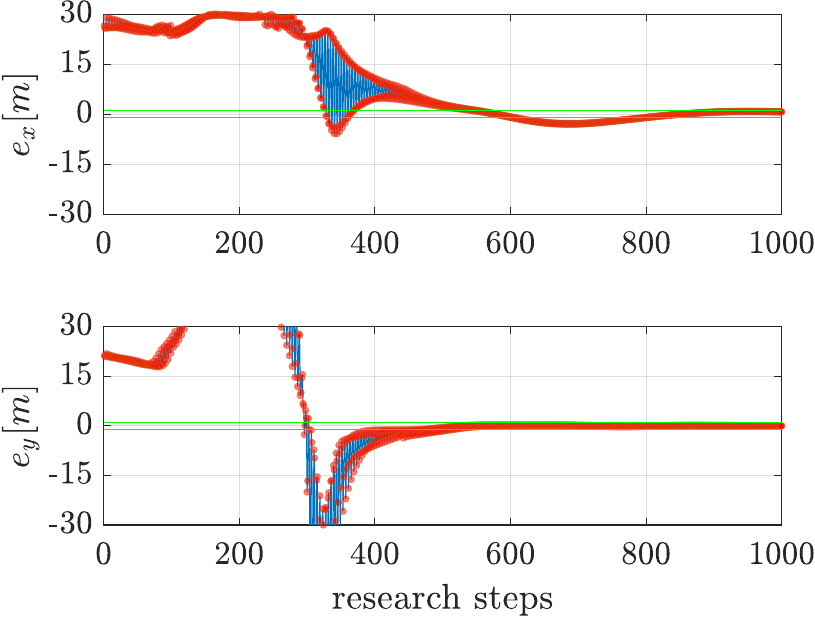}
        \caption{LS}
        \label{fig:LS}
    \end{subfigure}    
    \caption{Formation controllers comparison: combined ES+LS (a) vs ES and LS only (b-c). The top row shows the trajectories for the RX~agents, the bottom row reports the errors in the TX position estimation - solid green lines mark $\pm1[m]$ range.}
    \label{fig:comparison}
\end{figure*}

\subsection{Theoretical Working Principles of the Proposed Solution}

    We conceive our algorithm as a double-time scale system in which the formation controller \eqref{eq:formation_controller} represents the slow system, while the composition of {the ES gradient estimator}~\eqref{eq:ES} and the TX localiser~\eqref{eq:TXPosEstim} is the {fast system}. The time scale separation is induced by small $\gamma$. Then, as common in the singular perturbation theory \cite{kokotovic1999singular}, we define the so-called \textit{boundary layer} as the fast system evaluated at {$\gamma=0$}. We denote with $\hat{\nabla}_{i, \infty}(t)$ and $\hat{p}_{T, \infty}(t)$ as the asymptotic solutions of the boundary layer. Let $\sigma_{\infty}(t)$, $\hat{b}_{\text{ES}, \infty}(t)$, and $\hat{b}_{\text{LS}, \infty}({t})$ be \eqref{eq:sigma} and \eqref{eq:mean_agents} evaluated at $\hat{\nabla}_{i, \infty}(t)$ and $\hat{p}_{T, \infty}({t})$. Then, the so-called \textit{reduced system} is represented by the formation controller evaluated at $\hat{b}_{\text{ES}, \infty}(t)$,  $\hat{b}_{\text{LS}, \infty}(t)$, and $\sigma_{\infty}(t)$. We define $p_{c,\text{red}}(t)$ the solutions of the reduced system.

The formation design relies on the principle that the ARTVA SNR increases when the agents are closer to the TX. Let $\nabla_{i}^\star(p_i)$ be the gradient of $y(\cdot,0)$ evaluated at $p_i$ at zero noise. Then, for any given desired accuracy $c^\prime> 0$ and any noise level $\overline{\nu} > 0$ there exists a maximum radius $\rho^\prime$ such that $\|\nabla_i(p_i)-\nabla_i^\star(p_i)\| \le c^\prime$ for all $\|p_i-p_T\| \le \rho^\prime$. Moreover, the gradient estimation error $\|\hat{\nabla}_{i,\infty}-\nabla_i\|$ can be tuned via standard averaging arguments \cite{sanders2007averaging}. In detail, for any $\omega \le 2\pi/3$, $\kappa \in (0,1)$, 
$c^{\prime\prime} > 0$ there exist $\overline{\alpha}> 0$, $\overline{\delta}> 0$, $\rho^{\prime\prime} > 0$ such that $\|\hat{\nabla}_{i,\infty}-\nabla_i\| \le c^{\prime\prime}$ for all $\alpha \in (0, \overline{\alpha})$, $\delta \in (0, \overline{\delta})$, and $\|p_i-p_T\| \le\ \rho^{\prime\prime}$. {This implies that the formation radius $r:= \max_{i,j \in \{1,\dots,n\}} \|p_i-p_j\|$ must be chosen as $r < \rho:=\min\{\rho^{\prime},\rho^{\prime\prime}\}$.} Hence, there exists a set of initial conditions, namely $\mathcal{P}_0$, such that $p_i \in p_T + \mathcal{B}_{\rho}$.

Given $y$ as in \eqref{eq:y}, 
there exists $c   \in (0,1)$ ($c \approx 0.9$ ) such that
$c \le \langle \nabla_i^\star(p_i)/\|\nabla_i^\star(p_i)\|, (p_T-p_i)/\|p_T-p_i\| \rangle \le 1$. This implies that we can design our system to keep a tight coherency between $\hat{b}_{\text{ES}, \infty}(t)$ and $(p_T-p_i(t))/\|p_T-p_i(t)\|$, for all $t \in \N$, thanks to the previous arguments about the estimation accuracy of the nominal gradient $\nabla_i^\star(p_i(t))$. Let $b_{i,\infty}(t)$ be defined as \eqref{eq:Bearing} evaluated at $\hat{\nabla}_{i, \infty}(t)$. Since, the boundary layer TX estimation problem, \textit{i.e.}, problem \eqref{eq:hat_pt_star} evaluated at $b_i(t) = b_{i,\infty}(t)$, becomes locally convex around $p_T$, 
then we can demonstrate that for any desired estimation performance index $d > 0$ there exists $\underline{N} > 0$ such that $\|\hat{p}_T(t,N)-p_T\| \le d$ for all $N \ge \underline{N}$~\cite{boyd2004convex}. Since also the boundary layer TX estimator, \textit{i.e.} \eqref{eq:TXPosEstim} evaluated at $b_i(t) = b_{i,\infty}(t)$, works well if all $b_{i,\infty}(t)$ are coherent with $(p_T-p_i(t))/\|p_T-p_i(t)\|$, we have that $\langle \hat{b}_{\text{ES},\infty}(t), \hat{b}_{\text{LS},\infty}(t)\rangle > 0$, when the formation is far from $p_T$. This implies that the reduced system steers the formation in the right direction when far from the TX. On one hand, $\|\hat{b}_{\text{ES}, \infty}\|$ reduces as $p_c$ approaches the target. On the other, the estimation $\hat{p}_T$ gets more reliable when ${\bar{p}}_c$ approaches $p_T$ because the formation, {whose agents are } supposed uniformly distributed on a sphere of radius $r$, surrounds $p_T$. This 
makes $p_{c,\text{red}}(t)$ converge in a neighbourhood of $p_T$.

Finally, once the boundary layer and the reduced system have been assessed to have an asymptotic stable behavior, standard singular perturbation arguments guarantee the existence of $\overline{\gamma} > 0$ such that also the original system \eqref{eq:y}-\eqref{eq:formation_controller} is asymptotically stable for any ${\gamma} \in (0, \overline{\gamma})$. 

The performance required in 2) of Problem~\ref{prob} is verified through a suitable selection of the design parameters, including $\delta$ and $p_{ib}$ for $i=1,\ldots, n$.

\section{Simulation Results}

We assess the performance of the proposed solution through numerical simulations, {focusing on} the 2D case. Specifically, we consider a group of $n = 4$ RX agents placed at the corner of a square formation, whose edge is $\ell = 10$m long, initially centered in $\bar{p}_c({1})=\col(25,25)${m} and
required to localise the position $p_T = \col(0,0)${m}.
In implementing the outlined solution, we rest on the selection $\Pi=1$ W  and $\nu_i=10^{-5} $T for the sensing model~\eqref{eq:y} and $N=200$ for the period in~\eqref{eq:time}. The ES scheme~\eqref{eq:ES} is run by setting $\omega = \pi/2$ and $\alpha = 10^{-2}$ and  imposing $\hat{\nabla}_i({1}) \sim \mathcal{N}(0,10^{-5}I_3)$. In the LS-based estimator, we choose $\hat{\rho} = 50$ m because it represents a typical ARTVA range. Finally, the fleet is governed by imposing {$\gamma=10^{-2}$}. 

{Figure}~\ref{fig:comparison} reports the overall behavior of the RX~agents in localising and surrounding the TX, showing in the top panels their trajectories (black markers) and the trend of the TX position estimation (red dots). {The mid and bottom panels report the trend of the components of the vector $e = \hat{p}_T-p_T$ at the end of each research step which coincides with the update of the fleet's centroid. The superior performance of the newly designed ES+LS formation control solution ({Figure}~\ref{fig:ES+LS}) can be appreciated in terms of convergence {speed} (with respect to the ES-only formation control, obtained by forcing the parameter ${\sigma(t,N)} $ in~\eqref{eq:sigma} to 1 and depicted in {Figure}~\ref{fig:ES}) and the smoothness of the trajectories (with respect to the LS-only formation control, obtained by forcing ${\sigma(t,N)} $ to 0 and depicted in {Figure}~\ref{fig:LS}). This result is also confirmed by looking at the TX position estimation errors of the bottom row, which highlight the effective trend of the combined solution (a), the slow convergence of the ES-only regulator (b), and the more erratic behavior of the LS-only. 

In {Figure}~\ref{fig:ES+LS_sigma}, we report the value of the parameter ${\sigma(t,N)} $ computed according to~\eqref{eq:sigma}, referring to the combined algorithm ES+LS updating the RX~agents' position as in~\eqref{eq:bar_pc_plus}. 
{As expected}, the algorithm {trusts} the ES direction $\hat{b}_{\text{ES}}$ (${\sigma(t,N)} >0.5$), being this more {reliable than $\hat{b}_{\text{LS}}$ during the first phase of the research. C}onversely, as long as the RX~agents get closer to the TX, their LS direction becomes more reliable and the corresponding contribution allows to boost the localisation performance. 

{Then, we compare} the performance of the proposed solution {in terms of} the  distance between the {formation's centroid} $\bar{p}_c$ and the actual {TX position} $p_T$, namely $d\left(\bar{p}_c,p_T\right)=\Vert \bar{p}_c-p_T \Vert$, reported in {Figure}~\ref{fig:ES+LS_distance}. {It appears that the ES+LS solution (red) behaves like the ES-only scheme (green) in the initial transient, and it behaves like the LS-only paradigm (blue) asymptotically. Then, the benefits of the combined solution ES+LS stay in the improved transient.}

\begin{figure}[t!]
    \centering
    \includegraphics[width=0.78\columnwidth, trim ={1pt 1pt 1pt 1pt},clip]{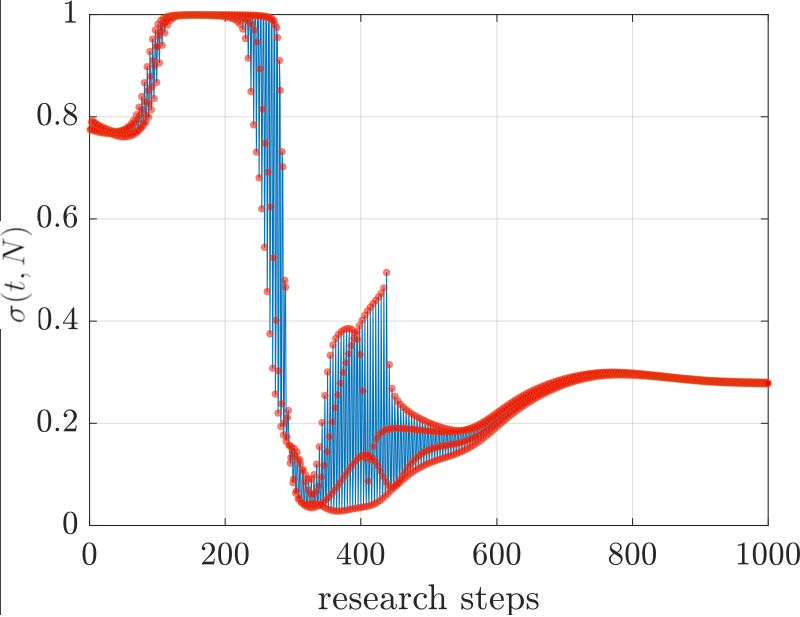}
    \caption{Trend of ${\sigma(t,N)}$  depicted in terms of punctual values for each research step (red dots) and interpolated dynamics (blue line). In the convex combination \eqref{eq:bar_pc_plus}, ${\sigma(t,N)}=1$ implies that the fleet position is updated according the ES output, while ${\sigma(t,N)} \rightarrow 0$ indicates the formation controller major reliance on LS position estimation.}
    
    \label{fig:ES+LS_sigma}
\end{figure}

\begin{figure}[t!]
    \centering
    \includegraphics[width=0.78\columnwidth]{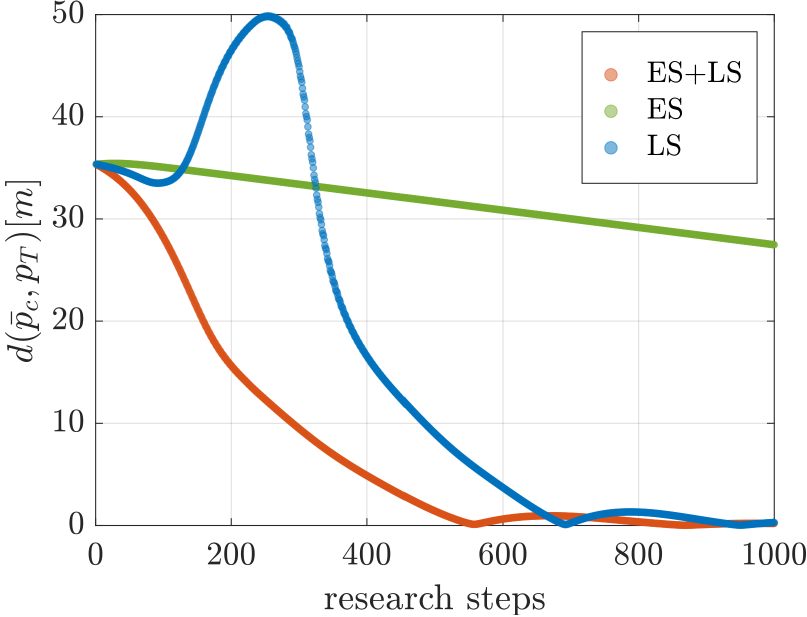}
    \caption{Evolution of the distance $d(\bar{p}_c,p_T)$ between the formation centroid $\bar{p}_c$ and the (true) TX position $p_T$.}
    \label{fig:ES+LS_distance}
\end{figure}

To conclude, we compare the performance of the ES+LS approach with the ES-only and LS-only approaches within a statistical framework. We conducted 1000 Monte Carlo runs, maintaining the same parameters as in previous tests. Fixing a maximum of $10^4$ research steps, the focus is on the number of those required to ensure that {$\|e\| \le 2$ meters} for $30$ consecutive steps.
Figure~\ref{fig:boxplot} shows that the mean number of steps is $4,287 \cdot 10^3$ for the ES-only solution and $6,696 \cdot 10^2$ for the LS-only one while adopting the combined ES+LS approach yields a mean of $4,411\cdot 10^2$ steps (green dots).
The LS-only and ES+LS solutions exhibit similar mean performance; however, the LS-only method presents a higher number of outliers and greater variance. This observation aligns with the fact that the LS-based estimator heavily relies on reliable input data. Consequently, this concluding comparison reaffirms the effectiveness of the ES+LS approach in speeding up the localisation process, particularly in comparison to the ES-only method. 
Moreover, it also highlights the superior robustness of the ES+LS approach when compared with the LS-only method.

\section{Conclusions and Further Developments}

In this paper, we outline a novel solution to coordinate a multi-agent system involved in the localisation of a target consisting of an electromagnetic source. The proposed approach aims to combine the reliable features of the ES paradigm with the promptness of cooperative LS estimation. Any agent utilizes the output of a private ES-based ARTVA's gradient estimator method to infer a reliable bearing towards the unknown target. The set of bearings gathered from the multi-agent system then serves as input to a centralized algorithm responsible for estimating the target's position by recursively solving an LS minimization problem. Finally, the formation controller is designed to steer the system's centroid by properly accounting for both the ES and LS contributions, to reduce the localisation time. The results of the numerical simulations assess the effectiveness of the proposed approach. Compared to the more traditional ES-only approach, the new solution achieves an order of magnitude reduction (from $10^3$ to $10^2$) in the number of search steps and, consequently, in the time required to confidently localise the target within a $2$-meter accuracy threshold. In addition, it turns out to be more robust with respect to the LS-only approach. 
Further research steps include the formal proof of the asymptotic stability of the proposed algorithm and the extension to the localisation of multiple targets.

\begin{figure}[t!]
    \centering
   \includegraphics[width=0.78\columnwidth]{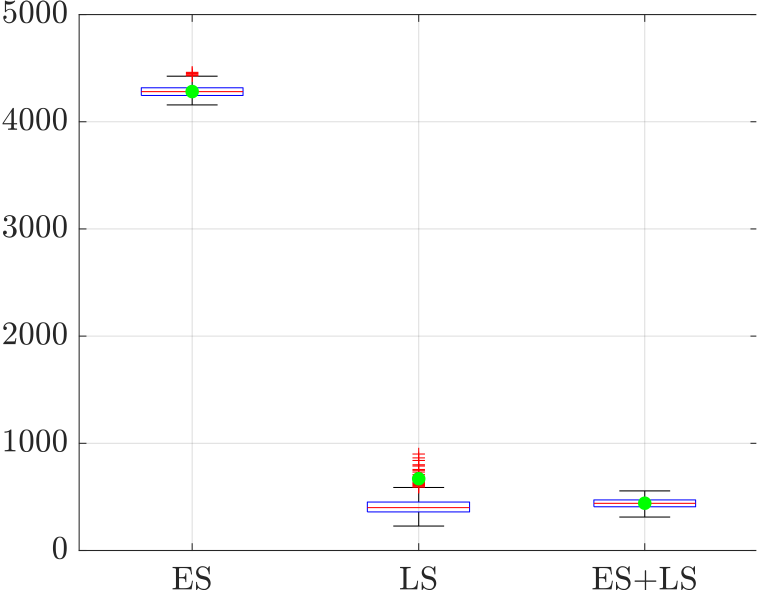}
    \caption{Statistical analysis of the number of research steps required to ensure that {$\|e\| \le 2$ meters} for $30$ consecutive steps - green dots indicate the mean number.}
    \label{fig:boxplot}
\end{figure}

\bibliographystyle{ieeetr}
\bibliography{bibfile}


\end{document}